\documentclass[12pt, journal, onecolumn]{IEEEtran}
\usepackage{amsmath}
\usepackage{amssymb}
\usepackage{cite}
\usepackage{epsfig}
\usepackage{epsf}
\usepackage{theorem}

\newtheorem{thm}{Theorem}
\newtheorem{lem}{Lemma}

\begin{document}
\pagenumbering{arabic}
\title{\textbf{Scheduling and Codeword Length Optimization in Time Varying Wireless Networks}}
\author{Mehdi Ansari Sadrabadi, Alireza Bayesteh and Amir K. Khandani \\
\small Coding \& Signal Transmission Laboratory(www.cst.uwaterloo.ca)\\
Dept. of Elec. and Comp. Eng., University of Waterloo\\ Waterloo, ON, Canada, N2L 3G1 \\
Tel: 519-884-8552, Fax: 519-888-4338\\e-mail: \{mehdi, alireza,
khandani\}@cst.uwaterloo.ca} \maketitle

\begin{abstract}
In this paper, a downlink scenario in which a single-antenna base
station communicates with $K$ single antenna users, over a
time-correlated fading channel, is considered.  It is assumed that
channel state information is perfectly known at each receiver, while
the statistical characteristics of the fading process and the fading
gain at the beginning of each frame are known to the transmitter. By
evaluating the random coding error exponent of the time-correlated
fading channel, it is  shown that there is an optimal codeword length
which maximizes the throughput. The throughput of
the conventional scheduling that transmits to the user with the maximum
signal to noise ratio is examined using both fixed length codewords and variable
length codewords. Although optimizing the codeword length improves
the performance, it is shown that using the conventional scheduling,
a gap of $\Omega(\sqrt{\log \log\log K})$ exists between the
achievable throughput and the maximum possible throughput of the
system.  A simple scheduling that considers both the
signal to noise ratio and the channel time variation is proposed.  It is shown that by using this scheduling, the gap between the achievable throughput and
the maximum throughput of the system approaches zero.
\end{abstract}


\section{Introduction}
In wireless networks, diversity is a means to combat the time
varying nature of the communication link. Conventional diversity techniques over point-to-point links, such
as spatial diversity and frequency diversity are widely used and  offer performance improvements. In multiuser wireless
systems, there exists another form of diversity, called
\textit{multiuser diversity}~\cite{dumbantenna}. In a broadcast
channel where users have independent fading and feed back their
signal to noise ratio (SNR) to the base station (BS), system
throughput is maximized by transmitting to the user with the
strongest SNR~\cite{dumbantenna, hanly}.

Multiuser diversity was introduced first by Knopp and Humblet
\cite{KH}. It is shown that the optimal transmission strategy in the
uplink of multiuser systems using power control is to only let the
user with the largest SNR transmit. A similar result is shown to be
valid for the downlink~\cite{Tse-mud}. Multiuser diversity underlies
much of the recent works for downlink scheduling \cite{agrawal,
liu, liu1,  borst} as in
Qualcomm's high data rate (HDR) system~\cite{bender,cdma}.
In~\cite{jalali, borst},  the opportunistic scheduling is based on the
highest data rate which can be reliably transmitted to each user. Distributed scheduling is proposed
in an uplink scenario, where full channel state information (CSI) is
not required at the transmitter~\cite{qin, shamaitel}. Multiuser diversity has also been studied in the context of multiple antenna systems~\cite{dumbantenna}, \cite{multiuser-mimo} and  ad-hoc
networks~\cite{Tse-mobi}.

In wireless networks, the rate of channel variations is
characterized by maximum Doppler frequency which is proportional to
the  velocity. Utilizing multiuser diversity in such  environments
needs to be revisited since the throughput depends not only on the
received SNR, but also on how fast the channel varies over time.

In this paper, we consider a broadcast channel in which a BS
transmits data to a large number of users in a time-correlated flat
fading environment. It is assumed that CSI is perfectly known to the receivers, while BS only knows the
statistical characteristics of the fading process for all the users
(which is assumed to be constant during a long period). Moreover,
each user feeds back its channel gain to the BS at the beginning of
each frame.  Based on this information, BS selects a single user for
transmission in each frame, in order to maximize the throughput. For
the case of Additive White Gaussian Noise (AWGN) or block fading, it
is well known that increasing the codeword length results in
improving the achievable throughput. However, in a time varying
channel, it is not possible to obtain arbitrary small error
probabilities by increasing the codeword length. In fact, increasing
the codeword length also results in increasing the fading
fluctuations over the frame, and consequently, the throughput will
decrease. Therefore, it is of interest to find the optimum codeword
length which maximizes the throughput.

In this paper,  a downlink scenario in which a single-antenna base
station communicates with $K$ single antenna users, over a
time-correlated fading channel, is considered. We analyze different
user selection strategies; i)~the BS transmits data to the user with
the strongest SNR using fixed length codewords (conventional
multiuser scheduling), ii)~the BS transmits data to the user with
the strongest SNR using variable length codewords, and iii)~the BS
transmits data to the user that achieves the maximum throughput
using variable length codewords. We show that in all cases the
achievable throughput scales as $\log \log K$. Moreover, in cases
(i) and (ii), the gap between the achievable throughput and the
maximum throughput scales as $\sqrt{\log \log \log K}$, while in
case (iii), this gap approaches zero.

The rest of the paper is organized as follows. In Section
\ref{model}, the model of time-correlated fading channel is
described.  In Section \ref{analysis},  different user selection
strategies are discussed and  the corresponding throughput of the
system is derived for each strategy, for $K \to \infty$. Finally, in Section \ref{conclusion}, we conclude the
paper.

Throughout this paper, $\mathbb{E} \{ .\}$ and $\mbox{var} \{
.\}$ represents the expectation and variance, respectively, ``$\log
$" is used for the natural logarithm, and rate is expressed in
\emph{nats}. For given functions $f(N)$ and $g(N)$, $f(N)=O(g(N))$ is
equivalent to $\lim_{N \rightarrow \infty} \left| \frac{f(N)}{g(N)}
\right| < \infty$, $f(N)=o(g(N))$ is equivalent to $\lim_{N
\rightarrow \infty} \left| \frac{f(N)}{g(N)} \right| =0$,
$f(N)=\omega(g(N))$ is equivalent to $\lim_{N \rightarrow \infty}
\frac{f(N)}{g(N)} = \infty$, and $f(N)=\Omega(g(N))$ is equivalent
to $\lim_{N \rightarrow \infty} \frac{f(N)}{g(N)} =c$, where
$0<c<\infty$.

\section{System model}\label{model}
The channel of any given  user is modeled as
a time-correlated fading process.  It is assumed that the channel
gain is constant over each channel use (symbol) and varies from
symbol to symbol, following a Markovian random process. Assume that the
fading gain of $k^{\rm{th}}$ user is $\boldsymbol
{h}_k=[h_{1,k},\ldots,h_{N_k,k}]^T$ where $ h_{i,k}, 1\leq i \leq
N_k$  are complex Gaussian random variables with zero mean and unit
variance and $N_k$ is the codeword length of the $k^{\rm{th}}$ user. The received
signal for the $k^{\rm{th}}$ user is given by
\begin{equation}
\boldsymbol{r}_k = \boldsymbol{S}_k\boldsymbol{h}_k +
\boldsymbol{n}_k,
\end{equation}
where  $\boldsymbol{S}_k = \mbox{diag}(s_{1,k}, s_{2,k}, \ldots,
s_{N_k,k})$ is the transmitted codeword with the power constraint\footnote{Obviously,
for maximizing the throughput, the power constraint translates to $ \mathbb{E} \{|s_{i,k}|^2 \} = P$.}
$ \mathbb{E} \{|s_{i,k}|^2 \} \leq P$, and
$\boldsymbol{n}_k$ is AWGN with zero mean and covariance matrix $\boldsymbol{I}$.
Assume that $h_{0,k}$ is the fading
gain at the time instant before $\boldsymbol{S}_k$ is transmitted.
The sequence $u_{i,k}=|h_{i,k}|$, $0 \leq i \leq N_k$, is assumed
to be a stationary ergodic chain with the following probability
density function \cite{fading}:
\begin{eqnarray}\label{pu0}
f_{u_{0,k}}(u)=\left\{ \begin{array}{ll} 2ue^{-u^2} & u\geq 0\\
0 & \textrm{otherwise}\end{array} \right.,
\end{eqnarray}
\begin{equation}\label{corrfading}
f(u_{1,k}, u_{2,k}, \cdots, u_{N_k,k} |
u_{0,k})=\prod_{i=1}^{N_k}q_k(u_{i,k} |u_{i-1,k}),
\end{equation}
where,
\begin{eqnarray}
q_k(u|v)=\left\{ \begin{array}{ll}
\frac{2u}{1-\alpha_k^2}\exp\left(-\frac{u^2+\alpha_k^2v^2}{1-\alpha_k^2}\right)\mathcal{I}_0(\frac{2\alpha_k
u v}{1-\alpha_k^2}) & u\geq 0 \\ 0 &  \textrm{otherwise}
\end{array} \right. \notag
\end{eqnarray}
in which $0 <\alpha_k <1$ describes the channel correlation
coefficient of the $k^{\rm{th}}$ user. It is assumed that $\alpha_k$, $1 \leq k \leq K$, are i.i.d. random variables with uniform distribution which remain fixed during the entire transmission, and $\mathcal{I}_0(.)$ denotes the modified
Bessel function of order zero. It is assumed that CSI is perfectly known at each receiver,
while the statistical characteristics of the fading process and
$u_{0,k}$, $1 \leq k \leq K$  are known to the transmitter.

\section{Throughput Analysis} \label{analysis}
In this section, we derive the achievable throughput of the system
in the asymptotic case of $K \to \infty$.
 We define the $k^{\rm{th}}$ user's throughput per channel use, denoted by $T_k$,  as
\begin{eqnarray} \label{T}
T_k \triangleq R_k (1-p_e (k)),
\end{eqnarray}
where $R_k$ is the transmitted rate per channel use and $p_e(k)$ is
the frame error probability for this user. Using the concept of
random coding error exponent \cite{gallager}, $p_e (k)$ can be
upper-bounded as
\begin{eqnarray} \label{pe}
p_e (k)\leq \inf_{0 \leq \rho \leq 1} e^{-N(E_{k}(\rho)-\rho
R_k)}.
\end{eqnarray}
For simplicity of analysis, we use this upper-bound in evaluating
the throughput. This bound is tight for rates close to the capacity
as used in \cite{gallagerdelay, telgal, kaplan}.

 Assuming $s_{i,k}, 1\leq i \leq N_k$, are Gaussian
and i.i.d., it is shown that the random coding error exponent for the
$k^{\rm{th}}$ user, $E_{k}(\rho)$, is given by~\cite{kaplan},
\begin{equation}\label{ekro}
E_{k}(\rho)= -\frac{1}{N_k}\log \mathbb{E}_{\boldsymbol{u}_k}
\left \lbrace
\prod_{i=1}^{N_k}\left(\frac{1}{1+\frac{P}{1+\rho}u_{i,k}^2}\right)^{\rho }
\right \rbrace.
\end{equation}
where $\boldsymbol{u}_k=[u_{1,k},\ldots,u_{N_k,k}]$.

In the following, we assume that  $u_{0,k}\gg 1$. Since in
strategies introduced in this work, a user is selected if the
corresponding initial fading gain is maximum or above a certain
threshold, this assumption is valid when the number of users is
large.
\begin{thm}\label{throu}
For the channel model described in the previous section, and
assuming $u_{0,k}$ is known,  we have
\begin{equation}\label{errorex}
E_{k}(\rho) = \frac{1}{N_k}\sum_{i=1}^{N_k}
\rho\log\left(1+\frac{Pu_{0,k}^2\alpha_k^{2i}}{(1+\rho)}\right)+
O\left(\frac{1}{\sqrt{u_{0,k}}}\right)-O\left(e^{-u^2_{0,k}}\right).
\end{equation}
\end{thm}
\textbf{Proof}: See Appendix A.

Minimizing (\ref{pe}) is equivalent to maximizing $E_k(\rho)-\rho
R_k$. Noting (\ref{errorex}), we have
\begin{eqnarray}\label{errorex2}
E_k(\rho)-\rho R_k
&=&\frac{1}{N_k}\sum_{i=1}^{N_k}\rho\log\left(1+\frac{Pu_{0,k}^2\alpha_k^{2i}}{(1+\rho)}\right)-\rho
R_k+O\left(\frac{1}{\sqrt{u_{0,k}}}\right)-O\left(e^{-u^2_{0,k}}\right)\notag \\
&=&\frac{1}{N_k}\sum_{i=1}^{N_k}\rho\log\left(\frac{Pu_{0,k}^2\alpha_k^{2i}}{(1+\rho)}\right)-\rho
R_k+\frac{1}{N_k}\sum_{i=1}^{N_k}\rho\log\left(1+\frac{(1+\rho)}{Pu_{0,k}^2\alpha_k^{2i}}\right) \notag \\
&+& O\left(\frac{1}{\sqrt{u_{0,k}}}\right)-O\left(e^{-u^2_{0,k}}\right)\notag \\
&=&\frac{\rho}{N_k}\sum_{i=1}^{N_k}\left(\log\left(Pu_{0,k}^2\right)+2i\log(\alpha_k)-\log(1+\rho)\right)-\rho
R_k \notag \\
&+&\frac{1}{N_k}\sum_{i=1}^{N_k} O\left(\frac{1}{u_{0,k}^2}\right)+ O\left(\frac{1}{\sqrt{u_{0,k}}}\right)-O\left(e^{-u^2_{0,k}}\right)\notag \\
&=&\rho[\log(Pu_{0,k}^2)+(N_k+1)\log(\alpha_k)-\log(\rho+1)-R_k]\notag \\
&+& O\left(\frac{1}{\sqrt{u_{0,k}}}\right)-O\left(e^{-u^2_{0,k}}\right) 
\end{eqnarray}
It is easy to show that  $\rho_k^{\rm{opt}}$ which maximizes
(\ref{errorex2}) for large values of $u_{0,k}$ is
\begin{eqnarray}\label{roopt}
\begin{array}{rl}
\log(1+\rho_k^{\rm{opt}})+\frac{\rho_k^{\rm{opt}}}{1+\rho_k^{\rm{opt}}}=\beta_k,
& \beta_k<\log(2)+\frac{1}{2}\\
 \rho_k^{\rm{opt}}=1, & \beta_k \geq
\log(2)+\frac{1}{2}
\end{array}
\end{eqnarray}
where
\begin{equation}\label{betadef}
\beta_k =\log(Pu_{0,k}^2)+(N_k+1)\log(\alpha_k)-R_k.
\end{equation}
Using (\ref{T}), (\ref{pe}),  (\ref{errorex}) and (\ref{errorex2}),  we have
\begin{eqnarray}\label{tsim}
T_k &=&R_k\left[1-e^{-N_k\left(E_k(\rho_k^{\rm{opt}})-\rho_k^{\rm{opt}}R_k\right)}\right] \notag \\
&=&R_k \left[1-e^{-\rho_k^{\rm{opt}}N_k\left(\log(Pu_{0,k}^2)+
(N_k+1)\log(\alpha_k)-\log(\rho_k^{\rm{opt}}+1)- R_k\right)}\right].
\end{eqnarray}
It is easy to show that $T_k$ is a concave function of variables
$R_k$ and $N_k$, and the values of $R_k$ and $N_k$ which maximize
the throughput ($R_k^{\rm{opt}}$ and $N_k^{\rm{opt}}$) satisfy the
following equations\footnote{Note that we have relaxed the condition of $N_k$ being integer. However, since the optimizing $N_k$ tends to infinity as $K \to \infty$, this assumption does not affect the result.}:
\begin{equation}\label{Ropt}
R_k^{\rm{opt}}=\log(Pu_{0,k}^2)+(2N_k^{\rm{opt}}+1)\log(\alpha_k)-\log(\rho_k^{\rm{opt}}+1),
\end{equation}
\begin{equation}\label{Nopt}
N_k^{\rm{opt}}=\sqrt{\frac{\log\left(1+\rho_k^{\rm{opt}}N_k^{\rm{opt}}R_k^{\rm{opt}}\right)}{\rho_k^{\rm{opt}}\log(\alpha_k^{-1})}}.
\end{equation}
It follows that $N_k^{\rm{opt}} \to \infty$ and $R_k^{\rm{opt}} \to
\infty$ as $u_{0,k} \to \infty$. Using (\ref{Ropt}) and
(\ref{Nopt}),  (\ref{tsim}) can be re-written as follows:
\begin{eqnarray}\label{tsimsim}
T_k=\left(\log\left(\frac{Pu_{0,k}^2}{\rho_k^{\rm{opt}}+1}\right)+(2N_k^{\rm{opt}}+1)\log(\alpha_k)\right)\left(1-\frac{1}{1+\rho_k^{\rm{opt}}N_k^{\rm{opt}}R_k^{\rm{opt}}}\right)
\end{eqnarray}
Substituting  (\ref{Ropt}) in (\ref{betadef}), we have
\begin{equation}\label{beta}
\beta_k=N_k^{\rm{opt}}\log(\alpha_k^{-1})+\log(\rho_k^{\rm{opt}}+1).
\end{equation}
From (\ref{roopt}) and (\ref{beta}), it is concluded that
\begin{eqnarray}\label{roex}
\rho_k^{\rm{opt}}=\left\{\begin{array}{ll}
 \frac{N_k^{\rm{opt}}\log(\alpha_k^{-1})}{1-N_k^{\rm{opt}}\log(\alpha_k^{-1})} & \quad N_k^{\rm{opt}}\log(\alpha_k^{-1})< \frac{1}{2}    \\
1 & \quad N_k^{\rm{opt}}\log(\alpha_k^{-1}) \geq \frac{1}{2}
\end{array}\right.
\end{eqnarray}
 Noting (\ref{Ropt}) and
(\ref{roex}), for $\alpha_k=1$, we have $\rho_k^{\rm{opt}}=0$ and
$R_k^{\rm{opt}} = \log(Pu_{0,k}^2)$ which corresponds to the
capacity of a quasi-static fading channel (for large values of channel gain $u_{0,k}^2$).

In the following, we obtain the asymptotic throughput of the $k^{\rm{th}}$ user. Since there are two regions for $\rho_k^{\rm{opt}}$ as shown in $(\ref{roex})$, we calculate the closed form formula of the throughputs of these two cases separately.
\begin{itemize}
\item ($\rho_k^{\rm{opt}}=1$):  The corresponding asymptotic throughput is
obtained by substituting (\ref{Ropt}) and (\ref{Nopt}) in (\ref{tsim})
as follows:
\begin{eqnarray}\label{Tasym}
T_{k} & = & \log\left(\frac{Pu_{0,k}^2}{2}\right)- 2
\sqrt{\log(\alpha_k^{-1})\log\log\left(\frac{Pu_{0,k}^2}{2}\right)} \times  \nonumber \\
&{}&\left(1+O\left(\frac{\log\log\log(u_{0,k})}{\log\log(u_{0,k})}\right)\right).
\end{eqnarray}
From (\ref{Nopt}), the optimum codeword length scales as follows:
\begin{eqnarray} \label{noptscal}
N_k^{\rm{opt}} \sim \sqrt{\frac{\log \log (Pu_{0,k}^2)}{\log \alpha^{-1}}}.
\end{eqnarray}
Note that if $\alpha_k$ is fixed and $\alpha_k \neq 1$, then
$N_k^{\rm{opt}}\log(\alpha_k^{-1}) \geq \frac{1}{2}$ for large
values of $u_{0,k}$ and (\ref{Tasym}) is appliable.
\item ($\rho_k^{\rm{opt}}<1$):
We limit the calculation of the throughput to the assumption of $N_k^{\rm{opt}}\log(\alpha_k^{-1}) \sim o(1)$ to be able to derive the following closed form formula by  using (\ref{tsimsim}) and  (\ref{roex}):
\begin{eqnarray}\label{Tasym0}
T_{k} & = & \log(Pu_{0,k}^2)- 2
\sqrt[3]{\log(\alpha_k^{-1})\log\log\left(Pu_{0,k}^2\right)} \times  \nonumber \\
&{}&\left(1+O\left(\frac{\log\log\log(u_{0,k})}{\log\log(u_{0,k})}\right)\right)-o(1).
\end{eqnarray}
The corresponding optimum codeword length scales as follows:
\begin{eqnarray} \label{noptscal2}
N_k^{\rm{opt}} \sim \sqrt[3]{\frac{\log \log (Pu_{0,k}^2) }{(\log \alpha^{-1})^2}} .
\end{eqnarray}
As we will see later, this special case is sufficient to accomplish the calculations required in \ref{sub3}.
\end{itemize}
From the above equations, it is concluded that the throughput not
only depends on the initial fading gain, $u_{0,k}$, but also on the
fading correlation coefficient. Moreover, the throughput is an
increasing function of the channel correlation coefficient.

In the following, we introduce three scheduling strategies in order to
maximize the throughput; i) Traditional scheduling in which the
user with the largest channel gain is selected (SNR-based scheduling)
and the codeword length is assumed to be fixed. ii) SNR-based scheduling with optimized codeword length regarding the
channel condition of the selected user, and iii) Scheduling
which exploits both the channel gain and the channel correlation coefficient of the users. The
asymptotic throughput of the system is derived under each
strategy for $K \to \infty$.

\subsection{Strategy I: SNR-based scheduling with fixed codeword length}
The BS transmits to the user with the maximum initial fading gain.
The codeword length of all users is fixed, i.e.,  $N_1=N_2= \cdots =
N_K =N$. The codeword length $N$ is selected such that the throughput of the system is maximized. The BS adapts the data rate
to maximize the throughput of the selected user.
\begin{thm}
The asymptotic throughput of the system under Strategy I scales as
\begin{equation}
\overline{\mathcal{T}}_1 \sim \log\left(\frac{P\log
K}{2}\right)-2\sqrt{\mathbb{E}\{\log(\alpha^{-1})\}\log\log\log K},
\end{equation}
as $K \to \infty$.
\end{thm}

For simplicity of
notation, we define $\upsilon_k \triangleq u_{0,k}^2$. Let
$\upsilon= \max_{1\leq k\leq K}\upsilon_k $ and $\alpha$ be the
corresponding correlation coefficient of the selected user. Setting
the derivative of (\ref{tsim}) with respect to $R_k$ to zero, we
find the rate of the selected user and the corresponding throughput
in terms of $\upsilon$ and $\alpha$ as follows\footnote{We drop user
index of parameters $R$ and $\rho^{ \rm{opt}}$ for the selected
user.} :
\begin{eqnarray}\label{Ropt1}
R&=&\log\left(\frac{P\upsilon}{1+\rho^{
\rm{opt}}}\right)+(N+1)\log(\alpha)-\frac{\log(1+\rho^{
\rm{opt}}NR)}{\rho^{\rm{opt}}N},
\end{eqnarray}
\begin{eqnarray}\label{t1av}
\mathcal{T}_1(\upsilon,\alpha)&=&\left[\log\left(\frac{P\upsilon}{1+\rho^{
\rm{opt}}}\right)+(N+1)\log(\alpha) -\frac{\log(1+\rho^{
\rm{opt}}NR)}{\rho^{ \rm{opt}} N}\right] \times \notag\\
&& \left[1-\frac{1}{1+\rho^{ \rm{opt}}NR}\right],
\end{eqnarray}
where noting (\ref{roopt}), $\rho^{\rm{opt}}$ is determined as
follows:
\begin{eqnarray}\label{roopt1}
\begin{array}{rl}
\log(1+\rho^{\rm{opt}})+\frac{\rho^{\rm{opt}}}{1+\rho^{\rm{opt}}}=\beta,
& \beta<\log(2)+\frac{1}{2}\\
 \rho^{\rm{opt}}=1, & \beta \geq
\log(2)+\frac{1}{2}
\end{array}
\end{eqnarray}
Using (\ref{betadef}) and (\ref{Ropt1}), we have
\begin{equation}\label{beta1}
\beta=\log(1+\rho^{\rm{opt}})+\frac{\log(1+\rho^{\rm{opt}}NR)}{\rho^{
\rm{opt}}N}.
\end{equation}
Let us define  $R^*$ and event $\mathcal{A}$ as follows:
\begin{eqnarray}\label{Rsdef}
 R^* &\triangleq& \log (P \upsilon) + (N+1)
\log(\alpha)\notag \\
&\stackrel{(\ref{Ropt1}), (\ref{beta1})}{=}& R+\beta,
\end{eqnarray}
\begin{equation}\label{conda}
\mathcal{A}\equiv \{R^*  > \frac{1}{2} \log \log K\},
\end{equation}
 In the following, we derive upper-bounds for the
throughput of the system in terms of $R^*$ and  $\mathcal{A}$ which we use later in Lemma \ref{lem1} and Lemma \ref{lem2}.
\begin{eqnarray}
\overline{\mathcal{T}}_1 &=& \mathbb{E}
\{\mathcal{T}_1(\upsilon,\alpha)\}
\notag\\
&\stackrel{(\ref{Ropt1}), (\ref{t1av})}{\leq}& \mathbb{E} \{R\}\notag\\
 &=& \mathbb{E} \{R | \mathcal{A}\}\mbox{Pr} \{\mathcal{A}\} +
\mathbb{E}\{R|\mathcal{A}^C\} \mbox{Pr} \{\mathcal{A}^C\} \notag \\
&\stackrel{(\ref{Rsdef})}{=}& \left( \mathbb{E} \{R^* |
\mathcal{A}\} - \mathbb{E} \{\beta | \mathcal{A}\}\right) \mbox{Pr}
\{\mathcal{A}\} + \mathbb{E}
\{R|\mathcal{A}^C\} \mbox{Pr} \{\mathcal{A}^C\} \notag\\
\label{khaf1} &\leq& \mathbb{E} \{R^*\} -\mathbb{E} \{\beta |
\mathcal{A}\} \mbox{Pr} \{\mathcal{A}\} \\
\label{khaf2} &\leq&\mathbb{E} \{R^*\}
\end{eqnarray}
where (\ref{khaf1}) is derived by replacing $R$ with $R^*$, noting $R\leq R^*$, and 
$\mbox{Pr} \{\mathcal{A}\}$ can be computed as follows:
\begin{eqnarray}
\mbox{Pr} \{\mathcal{A}\} &=& \mbox{Pr} \{\log (P \upsilon) + (N+1)
\log \alpha > \frac{1}{2} \log \log K\} \notag\\
&=& 1- \mbox{Pr} \{\log (P \upsilon) + (N+1)
\log \alpha \leq \frac{1}{2} \log \log K\} \notag\\
&=& 1-\int_{0}^{\infty} \mbox{Pr} \left. \left \lbrace \log \alpha <
\frac{1}{N+1} \left(\frac{1}{2} \log \log K - \log (P x) \right)
\right| x
\right \rbrace f_{\upsilon}(x) d x \notag\\
&=& 1-\int_{0}^{\infty} F_{\alpha} \left(e^{\frac{1}{N+1}
\left(\frac{1}{2} \log \log K - \log (P x) \right)} \right)
f_{\upsilon}(x) d x,
\end{eqnarray}
where $f_y(.)$ and $F_y(.)$ are probability density function and
cumulative density function of random variable $y$, respectively.
Noting that $\alpha$ has a uniform distribution, we have
\begin{eqnarray} \label{pra}
\mbox{Pr} \{\mathcal{A}\} &=& 1-e^{\frac{\log \log K}{2(N+1)}}
\int_{0}^{\infty} e^{-\frac{\log (P x)}{N+1}} f_{\upsilon}(x)
dx \notag\\
&\sim& 1-e^{-\frac{\log \log K+ 2\log(P)}{2(N+1)}},
\end{eqnarray}
where  the second line follows from the fact that $\upsilon \sim
\log K$, with probability one \cite{sharif}.

According to $(\ref{roopt1})$, there are two regions for $\rho^{
\rm{opt}}$ of the selected user. To obtain the throughput of the
system, we upper-bound the throughput in these two regions in Lemma
\ref{lem1} and Lemma \ref{lem2}, respectively. Then, we derive a
lower-bound for the throughput of the system in Lemma \ref{lem3}.
\begin{lem}\label{lem1}
Assuming $\beta < \log(2)+\frac{1}{2}$, the throughput of Strategy I
is upper-bounded as follows:
\begin{eqnarray}\label{tlem1}
\overline{\mathcal{T}}_{11}  \lesssim
 \log (P \log K)-(\log(\log \log K -2\log(2))) \mathbb{E}
\{ \log (\alpha^{-1})\}\mbox{Pr}\{\mathcal{A}\}.
\end{eqnarray}
\end{lem}
\textbf{Proof:} Using (\ref{roopt1}) and (\ref{beta1}) and noting
$\beta < \log(2)+\frac{1}{2}$, we obtain
\begin{eqnarray}\label{nloginq}
(\rho^{\rm{opt}})^{-1} + (\rho^{\rm{opt}})^{-2} = \frac{N}{\log
(1+\rho^{\rm{opt}} N R)}.
\end{eqnarray}
Noting $\rho^{\rm{opt}} < 1$, it follows from (\ref{nloginq}) that
\begin{equation}\label{nlognr}
N > \log(1+\rho^{\rm{opt}} N R).
\end{equation}
Assuming $R$ is large enough, from (\ref{nloginq}), we have
\begin{equation}\label{ron2}
\frac{\rho^{\rm{opt}}NR}{\log(1+\rho^{\rm{opt}}N R)}>2R \Rightarrow
\rho^{\rm{opt}} N > 2.
\end{equation}
Using (\ref{nlognr}) and (\ref{ron2}), we can write 
\begin{eqnarray}\label{Nkhaf}
N &\stackrel{(\ref{nlognr})}{>}& \mathbb{E}\left\{\log(1+\rho^{\rm{opt}} N R)|\mathcal{A}\right\}\mbox{Pr} \{\mathcal{A}\} \notag \\
&\stackrel{(\ref{ron2})}{>}& \mathbb{E}\{(\log (1+2R))|\mathcal{A}\}\mbox{Pr}(\mathcal{A}) \notag \\
&\stackrel{a}{>}& (\log(\log \log K -2\log(2))) \mbox{Pr}\{\mathcal{A}\}
\end{eqnarray}
where ($a$) results from  the fact that conditioned on $\mathcal{A}$, we have $R=R^*-\beta>\frac{1}{2} \log
\log K-\frac{1}{2}-\log(2)$.
Noting that $\upsilon  \sim  \log K + O(\log
\log K)$ with probability one~\cite{sharif}, we can write the
throughput of the system as follows: {\setlength\arraycolsep{.5pt}
\begin{eqnarray} \label{t1case1}
\overline{\mathcal{T}}_{11} & \stackrel{(\ref{Rsdef}),(\ref{khaf2})}{\leq} &
\mathbb{E} \{ \log (P \upsilon) +
(N+1)\log(\alpha)\} \notag\\
&=& \mathbb{E} \{ \log (P\upsilon)\} -(N+1) \mathbb{E}\{\log(\alpha^{-1})\} \notag\\
&\stackrel{(\ref{Nkhaf})}{\lesssim}&  \log (P \log K)-(\log(\log \log K
-2\log(2))) \mathbb{E} \{ \log
\alpha^{-1}\}\mbox{Pr}\{\mathcal{A}\}.
\end{eqnarray}}
 \rightline{$\blacksquare$}

\begin{lem}\label{lem2}
Assuming $\beta \geq \log(2)+\frac{1}{2}$, the throughput of Strategy I is upper-bounded as follows:
\begin{eqnarray}\label{t1case2}
\overline{\mathcal{T}}_{12} \lesssim \log \left(\frac{P \log
K}{2}\right) - 2 \sqrt{\mathbb{E} \{\log \alpha^{-1}\}} \sqrt{\log
\log\log K}.
\end{eqnarray}
\end{lem}
\textbf{Proof:} Noting $\beta \geq \log(2)+ \frac{1}{2}$, from
(\ref{roopt1}), we have $\rho^{\rm{opt}}=1$. Hence, using (\ref{Rsdef}) and
(\ref{khaf1}) and noting $\upsilon \sim \log K$,  we can write
\begin{eqnarray} \label{tcase2}
\overline{\mathcal{T}}_{12} &\leq& \log \left(P \log K\right) -
(N+1) \mathbb{E} \{\log (\alpha^{-1}) \}- \mathbb{E} \{\beta |
\mathcal{A}\} \mbox{Pr}
\{\mathcal{A}\} \notag\\
&\stackrel{(\ref{beta1})}{\lesssim}& \log \left(P \log K\right) -
(N+1) \mathbb{E} \{\log (\alpha^{-1}) \}- \mathbb{E} \{\log(2)+
\frac{\log(1+NR)}{N}  | \mathcal{A}\} \mbox{Pr}
\{\mathcal{A}\} \notag\\
&\stackrel{a}{\lesssim}& \log \left(P \log K\right) - (N+1)
\mathbb{E}
\{\log (\alpha^{-1}) \} \notag \\
& & - \left[\frac{\log(\frac{1}{2} \log \log
K-\frac{\log(1+N\log(P\log K))}{N}-\log(2))}{N} +
\frac{\log N}{N}+\log(2)\right]\mbox{Pr} \{\mathcal{A}\}, \notag \\
\end{eqnarray}
where ($a$) follows from the fact that conditioned on $\mathcal{A}$,
we have
\begin{eqnarray}
R&=&R^*-\beta \notag \\
&>&\frac{1}{2} \log \log K-\frac{\log(1+NR)}{N}-\log(2) \notag \\
&>&\frac{1}{2} \log \log K-\frac{\log(1+N\log(P\log K))}{N}-\log(2),
\end{eqnarray}
The last line results from the fact that $R<\log(P\log K)$ which follows from (\ref{Ropt1}).
Substituting (\ref{pra}) in (\ref{tcase2}), and setting the
derivative of $\overline{\mathcal{T}}_{12}$ to zero with respect to
$N$,  we obtain
\begin{eqnarray} \label{nopt}
N^{\rm{opt}} \sim \sqrt{\frac{\log \log \log K}{\mathbb{E} \{\log
\alpha^{-1}\}}} [1+o(1)].
\end{eqnarray}
Substituting (\ref{pra}) and (\ref{nopt}) in (\ref{tcase2}), the result of the lemma follows.
\rightline{$\blacksquare$}

\begin{lem}\label{tkm}
Assume that $N$ is set as in (\ref{nopt}). Then, under condition
$\mathcal{A}$, we have $\rho^{\rm{opt}}=1$ as $K \to \infty$.
\end{lem}
\textbf{Proof:} If $\rho^{\rm{opt}}<1$, then using (\ref{roopt1}),
we have $\beta<\frac{1}{2}+\log(2)$. Noting (\ref{nlognr}) and
(\ref{ron2}), we can write
\begin{eqnarray}
N^{\rm{opt}}&\stackrel{(\ref{nlognr})}{>}&\log(1+\rho^{\rm{opt}}N^{\rm{opt}}R) \notag \\
&\stackrel{(\ref{ron2})}{>}&\log(1+2R) \notag \\
&\stackrel{(\ref{Rsdef})}{=}&\log(1+2(R^*-\beta)) \notag \\
&>&\log(2R^*-2\log(2)) \notag \\
&\stackrel{(\ref{conda})}{\sim}&\log\log\log K,
\end{eqnarray}
which contradicts (\ref{nopt}). \\
\rightline{$\blacksquare$}

\begin{lem}\label{lem3}
The throughput of Strategy I is lower-bounded as follows:
\begin{eqnarray}\label{tlem3}
\overline{\mathcal{T}}_1 \gtrsim
 \log \left(\frac{P \log K}{2}\right) - 2 \sqrt{\mathbb{E}
\{\log \alpha^{-1}\}\log\log\log K}.
\end{eqnarray}
\end{lem}
\textbf{Proof:} Choosing $N=N^{\rm{opt}}$ and subsequently replacing $\rho^{\rm{opt}}=1$ (Lemma \ref{tkm}), we compute $R$
in (\ref{Ropt1}),  under condition $\mathcal{A}$, as follows:
\begin{eqnarray}\label{rless}
R&=&R^*-\beta \notag \\
&>& \frac{1}{2}\log\log K - \log(2)- \frac{\log(1+N^{\rm{opt}}R)}{N^{\rm{opt}}} \notag \\
&>& \frac{1}{2}\log\log K - \log(2)- \frac{\log(1+N^{\rm{opt}}\log\log K)}{N^{\rm{opt}}} \notag \\
&\sim& \frac{1}{2}\log\log K - \sqrt{\mathbb{E} \{\log
\alpha^{-1}\}\log\log\log K)}
\end{eqnarray}
Using (\ref{t1av}) and (\ref{Rsdef}), we can lower-bound
$\overline{\mathcal{T}}_1$ as
\begin{eqnarray} \label{t1l}
\overline{\mathcal{T}}_1 &\geq& \overline{\mathcal{T}}_{1
|\mathcal{A}} \mbox{Pr}
\{\mathcal{A}\} \notag\\
&=& \mathbb{E} \left. \left \lbrace \left[R^*-\log(1+\rho^{
\rm{opt}}) -\frac{\log(1+\rho^{ \rm{opt}}N^{\rm{opt}}R)}{\rho^{\rm{opt}}
N^{\rm{opt}}}\right] \left[1-\frac{1}{1+\rho^{ \rm{opt}}N^{\rm{opt}}R}\right]\right|
\mathcal{A} \right \rbrace \times
\notag\\
&&\mbox{Pr} \{\mathcal{A}\},
\end{eqnarray}
where $\overline{\mathcal{T}}_{1 |\mathcal{A}}$ denotes the
throughput of the system conditioned on $\mathcal{A}$. Since $\mathbb{E}\{R^*\}=\mathbb{E}\{R^*|\mathcal{A}\}
\mbox{Pr} \{\mathcal{A}\} + \mathbb{E}\{R^*|\mathcal{A}^C\}
\mbox{Pr} \{\mathcal{A}^C\}$, and $\mathbb{E}\{R^*|\mathcal{A}\}>
\mathbb{E}\{R^*|\mathcal{A}^C\}$, it follows that
$\mathbb{E}\{R^*|\mathcal{A}\} \geq \mathbb{E}\{R^*\}$. Having this
fact, noting Lemma (\ref{tkm}) and using (\ref{pra}), (\ref{rless}) and (\ref{t1l}), we can write
\begin{eqnarray} \label{t1low}
\overline{\mathcal{T}}_1 &\geq&
\left(\mathbb{E}\{R^*\}-\log(2)-\mathbb{E} \left. \left \lbrace
\frac{\log (1+ N^{\rm{opt}} \log\left(\frac{P\log
K}{2}\right))}{N^{\rm{opt}}}\right| \mathcal{A} \right \rbrace
\right) \times \notag\\
&&\left( 1-\frac{1}{1+N^{\rm{opt}}(\frac{1}{2}\log\log K -
\sqrt{\mathbb{E} \{\log \alpha^{-1}\}\log\log\log K)}) }\right)
\left(
1-e^{-\frac{\log \log K}{2(N^{\rm{opt}}+1)}}\right) \notag\\
&\sim& \log \left(\frac{P \log K}{2}\right) - 2 \sqrt{\mathbb{E}
\{\log \alpha^{-1}\}\log\log\log K} .
\end{eqnarray}
\rightline{$\blacksquare$}

\textbf{The Proof of Theorem 2}:
Lemma (\ref{lem1}) and  Lemma (\ref{lem2}) provide upper-bounds on two complementary cases where $\rho^{\rm{opt}}$ of the selected user is either less than $1$ or equal to $1$ in (\ref{tlem1}) and (\ref{t1case2}), respectively. Lemma (\ref{lem3}) lower-bounds the throughput of the system as in (\ref{tlem3}). Comparing (\ref{tlem1}), (\ref{t1case2}) and (\ref{tlem3}), we conclude the result of the theorem.\\
\rightline{$\blacksquare$}
 \textit{Remark 1-} To prove Theorem $2$,
we utilize the distribution function of $\alpha$ to calculate
$\mbox{Pr}(\mathcal{A})$. The value of $\mbox{Pr}(\mathcal{A})$ is
used in (\ref{t1case1}), (\ref{tcase2}) and (\ref{t1low}). For
$\mbox{Pr}(\mathcal{A})=1-o\left(\frac{1}{\log \log K}\right)$,
(\ref{t1case1}), (\ref{tcase2}) and (\ref{t1low}) are valid. Therefore, the
assumption of uniform distribution for the correlation coefficients
can be relaxed if $\mbox{Pr}(\mathcal{A})=1-o\left(\frac{1}{\log
\log K}\right)$.

\subsection{Strategy II: SNR-based scheduling with adaptive codeword length}
In this scheme, the BS transmits to the user with the maximum
initial fading gain. The rate and codeword length are selected to
maximize the corresponding throughput.

\begin{thm}
Assuming $K \to \infty$, the asymptotic throughput of the system
under Strategy II scales as follows:
\begin{equation}
\overline{\mathcal{T}}_2 \sim \log\left(\frac{P\log
K}{2}\right)-2\mathbb{E}\{\sqrt{\log(\alpha^{-1})}\}\sqrt{\log\log\left(\frac{P\log
K}{2}\right)}.
\end{equation}
\end{thm}
\textbf{Proof}: The throughput of the system can be written as
\begin{eqnarray} \label{t2}
\overline{\mathcal{T}}_2 = \overline{\mathcal{T}}_{2 | \mathcal{B}}
\mbox{Pr} \{\mathcal{B}\}+ \overline{\mathcal{T}}_{2| \mathcal{B}^C}
\mbox{Pr} \{\mathcal{B}^C\},
\end{eqnarray}
where $\mathcal{B}$ represents the event that $\rho^{\rm{opt}}=1$,
$\overline{\mathcal{T}}_{2| \mathcal{B}}$ denotes the throughput
conditioned on $\mathcal{B}$, and $\overline{\mathcal{T}}_{2|
\mathcal{B}^C}$ is the throughput of the system conditioned on
$\mathcal{B}^C$, the complement of $\mathcal{B}$. Using
(\ref{Tasym}), we can write
\begin{eqnarray}
\overline{\mathcal{T}}_{2| \mathcal{B}} &=& \mathbb{E} \left. \left
\lbrace \log\left(\frac{P\upsilon}{2}\right)- 2
\sqrt{\log(\alpha^{-1})\log\log\left(\frac{P\upsilon}{2}\right)}
\left(1+
O\left(\frac{\log\log\log(\upsilon)}{\log\log(\upsilon)}\right)\right)
\right| \mathcal{B} \right \rbrace, \notag
\\
\end{eqnarray}
where $\upsilon = \max_{1\leq k\leq K} \upsilon_k$, and $\alpha$ is
the channel correlation coefficient of the selected user. Noting
that $\upsilon  \sim  \log K + O(\log \log K)$ with probability one,
and $\upsilon$ and $\alpha$ are independent, we have
\begin{eqnarray} \label{t2a}
 \overline{\mathcal{T}}_{2| \mathcal{B}} &\sim& \log \left( \frac{P \log K}{2}\right) - 2\mathbb{E}\left. \left \lbrace\sqrt{\log(\alpha^{-1})} \right|
 \mathcal{B}\right \rbrace
\sqrt{\log\log \left( \frac{P \log K}{2}\right)} \times \notag\\
& &\left(1+O\left(\frac{\log \log \log \log K}{\log \log \log
K}\right)\right).
\end{eqnarray}
Using  (\ref{roex}) and (\ref{noptscal}), we can write
\begin{eqnarray}\label{acondn}
\mathcal{B} &\equiv& N^{\rm{opt}} \log (\alpha^{-1}) \geq
\frac{1}{2} \notag\\
&\cong& \sqrt{\log (\alpha^{-1})} \sqrt{\log \log \log K} \geq
\frac{1}{2}.
\end{eqnarray}
Uniform distribution for  $\alpha$ results in exponential
distribution for $X \triangleq \log (\alpha^{-1})$, i.e., $f_{X} (x)
= e^{-x} u(x)$. Let us define $\epsilon \triangleq \frac{1}{4 \log
\log \log K}$. $\mbox{Pr}\{\mathcal{B}\}$ can be derived as follows:
\begin{eqnarray}\label{prb}
\mbox{Pr}\{\mathcal{B}\}&=&\mbox{Pr} \{ \log(\alpha^{-1}) \geq \epsilon\}\notag \\
&=& e^{-\epsilon}
\end{eqnarray}
Using (\ref{prb}), we have
\begin{eqnarray} \label{elogalphainva}
\mathbb{E}\left. \left \lbrace\sqrt{\log(\alpha^{-1})} \right|
\mathcal{B}\right \rbrace &=& \frac{\int_{\mathcal{B}} \sqrt{x} e^{-x} dx}{\mbox{Pr}
\{\mathcal{B}\}}\notag\\
&=& \frac{\int_{\epsilon}^{\infty} \sqrt{x} e^{-x} dx}{\mbox{Pr} \{\mathcal{B}\}}\notag\\
&\sim& \frac{\mathbb{E} \{\sqrt{\log(\alpha^{-1})}\} - \epsilon \sqrt{\epsilon}
e^{-\epsilon}}{e^{-\epsilon}} \notag\\
&\sim& \mathbb{E} \{\sqrt{\log(\alpha^{-1})}\} (1+O(\epsilon)).
\end{eqnarray}
Similarly, we can write
\begin{eqnarray}\label{elogbc}
\mathbb{E}\left. \left \lbrace\sqrt{\log(\alpha^{-1})} \right|
\mathcal{B}^C\right \rbrace &=&  \frac{\int_{0}^{\epsilon} \sqrt{x} e^{-x} dx}{\mbox{Pr} \{\mathcal{B}^C\}}\notag\\
&\sim& \frac{ \epsilon \sqrt{\epsilon}
e^{-\epsilon}}{1-e^{-\epsilon}} \notag\\
&\sim& O(\sqrt{\epsilon}).
\end{eqnarray}
Using (\ref{tsimsim}), $\overline{\mathcal{T}}_{2 | \mathcal{B}^C}$
can be written as
\begin{eqnarray} \label{t2ac}
\overline{\mathcal{T}}_{2|\mathcal{B}^C} &=& \mathbb{E} \left. \left
\lbrace \left(\log \left(\frac{P \upsilon}{1+\rho^{\rm{opt}}}
\right)-(2N^{\rm{opt}}+1)\log(\alpha^{-1})\right)\left(1-\frac{1}{1+\rho^{\rm{opt}}N^{\rm{opt}}R^{\rm{opt}}}\right)
\right| \mathcal{B}^C\right
\rbrace \notag\\
&\stackrel{a}{\gtrsim}& \mathbb{E} \left. \left \lbrace \left(\log
\left(\frac{P \upsilon}{1+\rho^{\rm{opt}}}
\right)-2\frac{\rho^{\rm{opt}}}{1+\rho^{\rm{opt}}}-\log(\alpha^{-1})\right)\left(1-\frac{1}{1+\rho^{\rm{opt}}N^{\rm{opt}}R^{\rm{opt}}}\right)
\right| \mathcal{B}^C\right \rbrace \notag \\
&\stackrel{b}{\gtrsim}& \mathbb{E} \left. \left \lbrace \left(\log
\left(\frac{P \upsilon}{1+\rho^{\rm{opt}}}
\right)-2\frac{\rho^{\rm{opt}}}{1+\rho^{\rm{opt}}}-\log(\alpha^{-1})\right)\left(1-\frac{1}{1+R^{\rm{opt}}}\right)
\right| \mathcal{B}^C\right
\rbrace \notag\\
&\gtrsim& \log \left(\frac{P \log
K}{2}\right)-\mathbb{E}\{\log(\alpha^{-1})| \mathcal{B}^C\}-2 \notag \\
&\stackrel{(\ref{elogbc})}{\gtrsim}&\log \left(\frac{P \log K}{2}\right)-2-O(\sqrt{\epsilon}),
\end{eqnarray}
where ($a$) follows from (\ref{roex}) which implies $N^{\rm{opt}}\log(\alpha^{-1})=\frac{\rho^{\rm{opt}}}{1+\rho^{\rm{opt}}}$ conditioned on $\mathcal{B}^C$ and ($b$) results from the following inequality:
\begin{eqnarray}\label{n2inq}
\rho^{\rm{opt}}N^{\rm{opt}}R^{\rm{opt}} &\stackrel{(\ref{roex})}{=}&
\frac{R^{\rm{opt}}(N^{\rm{opt}})^2\log(\alpha^{-1})}{1-N^{\rm{opt}}\log(\alpha^{-1})}\notag\\
&{\geq}& R^{\rm{opt}}(N^{\rm{opt}})^2\log(\alpha^{-1})\notag \\
&\stackrel{(\ref{Nopt})}{=}& R^{\rm{opt}}\frac{\log(1+\rho^{\rm{opt}}N^{\rm{opt}}R^{\rm{opt}})}{\rho^{\rm{opt}}}\notag \\
&\stackrel{a}{\geq}& R^{\rm{opt}} \log(1+N^{\rm{opt}}R^{\rm{opt}})\notag \\
&\gtrsim& R^{\rm{opt}}
\end{eqnarray}
where ($a$) follows from the fact that $\frac{\log(1+\rho^{\rm{opt}}N^{\rm{opt}}R^{\rm{opt}})}{\rho^{\rm{opt}}}$ is a decreasing function of $\rho^{\rm{opt}}$.
Moreover, using (\ref{tsimsim}) and noting
that $\upsilon  \sim  \log K$ with probability one, we have
\begin{eqnarray} \label{t2acu}
\overline{\mathcal{T}}_{2|\mathcal{B}^C} &\leq& \mathbb{E} \left.
\left \lbrace \log \left(P \upsilon
\right) \right | \mathcal{B}^C\right \rbrace \notag \\
&\lesssim&\log (P \log K)
\end{eqnarray}
Combining (\ref{t2ac}) and (\ref{t2acu}), we have
\begin{equation}\label{inqt2}
\log \left(\frac{P \log K}{2}\right)-2-O(\sqrt{\epsilon}) \lesssim
\overline{\mathcal{T}}_{2|\mathcal{B}^C} \lesssim \log \left(\frac{P
\log K}{2}\right)+ \log(2)
\end{equation}
Substituting (\ref{t2a}) and (\ref{inqt2}) in (\ref{t2}) and noting
(\ref{prb}) and (\ref{elogalphainva}), after some manipulations, we have
\begin{eqnarray}
\overline{\mathcal{T}}_2 & \sim & \left(\log \left( \frac{P \log
K}{2}\right) - 2\mathbb{E} \{\sqrt{\log(\alpha^{-1})}|\mathcal{B}\} \sqrt{\log
\log\left( \frac{P \log K}{2}\right)} \right) \mbox{Pr} \{\mathcal{B}\} \notag \\
&+& \left(\left(\frac{P \log K}{2}\right) + \Omega(1) \right)\mbox{Pr} \{\mathcal{B}^C\}\notag \\
&\sim& \log \left( \frac{P \log K}{2}\right) -2\mathbb{E} \{\sqrt{\log(\alpha^{-1})}\} \sqrt{\log
\log\left( \frac{P \log K}{2}\right)} \notag \\
&+&2\mathbb{E} \{\sqrt{\log(\alpha^{-1})}\} \sqrt{\log
\log\left( \frac{P \log K}{2}\right)}\epsilon \sqrt{\epsilon}e^{-\epsilon} +\Omega(1)O(\sqrt{\epsilon})\notag \\
&\sim& \log \left( \frac{P \log K}{2}\right) - 2\mathbb{E} \{\sqrt{\log(\alpha^{-1})}\} \sqrt{\log
\log\left( \frac{P \log K}{2}\right)}+O(\sqrt{\epsilon})
\end{eqnarray}
which completes the proof of Theorem 3.\\
\rightline{$\blacksquare$} \textit{Remark 1-} To prove Theorem 3, we
used the following properties:
\begin{eqnarray}
\mathbb{E}\left. \left \lbrace\sqrt{\log(\alpha^{-1})}
\right|\mathcal{B}\right \rbrace \sim \mathbb{E}
\{\sqrt{\log(\alpha^{-1})}\} (1+O(\epsilon))\notag\\
\mathbb{E}\left. \left \lbrace\sqrt{\log(\alpha^{-1})}
\right|\mathcal{B^C}\right \rbrace \sim O(\sqrt{\epsilon}) \notag \\
\mbox{Pr}\{\mathcal{B}^C\}\sim O(\epsilon),
\end{eqnarray}
where $\epsilon \sim \frac{1}{\log \log
\log K}$. The theorem is valid for any distribution function of $\alpha$ that satisfies the above properties.   \\
\textit{Remark 2-} Since $\mathbb{E} \{\sqrt{x} \} \leq \sqrt{\mathbb{E} \{x\}}$,
for $x >0$, it is concluded that the achievable rate of Strategy II
is higher than that of Strategy I. More precisely,
\begin{eqnarray}
\overline{\mathcal{T}}_2 -\overline{\mathcal{T}}_1  &\sim& 2 \left(
\sqrt{\mathbb{E} \{\log(\alpha^{-1})\}} - \mathbb{E}
\{\sqrt{\log(\alpha^{-1})} \} \right) \sqrt{\log \log \log K}.
\end{eqnarray}
For the case of uniform distribution for $\alpha$, we have
\begin{eqnarray}
\overline{\mathcal{T}}_2 -\overline{\mathcal{T}}_1 &\sim& 0.228
\sqrt{\log \log \log K}.
\end{eqnarray}
\textit{Remark 3-} Although $\lim_{K \to \infty}
\frac{\overline{\mathcal{T}}_1}{\overline{\mathcal{T}}_{\max}} =
\lim_{K \to \infty}
\frac{\overline{\mathcal{T}}_2}{\overline{\mathcal{T}}_{\max}} =1$,
where $\overline{\mathcal{T}}_{\max} \sim \log \left(P \log
K\right)$ is the maximum achievable throughput for a quasi-static
fading channel \cite{sharif}, there exists a gap of
$\Omega(\sqrt{\log \log \log K})$ between the achievable throughput
of Strategies I and II, and the maximum throughput. As we show
later, this gap is due to the fact that the channel correlation
coefficients of the users are not considered in the scheduling. In
fact, this gap approaches zero by exploiting the channel
correlation, which is discussed in Strategy III.

\subsection{Strategy III: Scheduling based on both SNR and channel
correlation coefficient with adaptive codeword length }\label{sub3}
To maximize the throughput of the system, the
user which maximizes the expression in (\ref{tsimsim}) should be serviced.
Here, for simplicity of analysis,
 we propose a  sub-optimum scheduling that considers
the effect of both SNR and channel correlation in the user
selection. In this strategy, each user is required to feed back its
initial fading gain only if it is greater than a pre-determined
threshold $\sqrt{\Theta}$, where $\Theta$ is a function of the
number of users. Among these users, the BS selects  the one with the
maximum channel correlation coefficient. The data rate and codeword
length are selected to maximize the corresponding throughput. The
following theorem gives the system throughput under this strategy.
\begin{thm}\label{s3}
 Using Strategy III, with $\Theta$ satisfying
\begin{equation}\label{tetta}
\log K -o(\log K) \lesssim  \Theta \lesssim \log K - \log\log K -\omega(1),
 \end{equation}
the throughput of the system scales as
\begin{eqnarray}\label{T3}
\overline{\mathcal{T}}_3 &\gtrsim&  \log\left(P\log K\right)- o(1)
\end{eqnarray}
\end{thm}
\textbf{Proof:} Define ${\cal S} \triangleq \{k|\upsilon_k \geq
\Theta\}$ and $\alpha_{\max}\triangleq \max_{k\in {\cal S}}\alpha_k
$. Let $\upsilon$ be the squared initial fading gain of the user
corresponding to $\alpha_{max}$. We define the event $\mathcal{G}$
as follows:
\begin{equation}\label{gevent}
\mathcal{G}\equiv N^{\rm{opt}}\log(\alpha_{\max}^{-1})\sim
o(1),
\end{equation}
where $N^{\rm{opt}}$ is the corresponding codeword length as
computed from (\ref{Nopt}). Using (\ref{Tasym0}) and (\ref{gevent}),
we can write
\begin{equation}\label{t31}
\overline{\mathcal{T}}_3\geq
\mbox{Pr}\{\mathcal{G}\}\mathbb{E}\{\mathcal{T}_3(\upsilon,\alpha_{\max})|\mathcal{G}\}
\end{equation}
where following (\ref{Tasym0}),
\begin{eqnarray}\label{t3con}
\mathbb{E}\{\mathcal{T}_3(\upsilon,\alpha_{\max})|\mathcal{G}\} &=&
\mathbb{E} \left \lbrace \log\left(P\upsilon\right)
-2\sqrt[3]{\log(\alpha_{\max}^{-1})}{} \right. \nonumber \\ & &
\left. \sqrt[3]{\log\log\left(P\upsilon\right)} \left. \left[1+ O\left(\frac{\log\log\log(\upsilon)}{\log\log(\upsilon)}\right)\right]-o(1) \right |\mathcal{G}\right \rbrace 
\end{eqnarray}
Noting that $\{\mathcal{T}_3(\upsilon,\alpha_{\max})|\mathcal{G}\}$ in (\ref{t3con})
is an increasing function of $\upsilon$, we have
\begin{eqnarray}\label{Teq}
\mathbb{E}\{\mathcal{T}_3(\upsilon,\alpha_{\max})|\mathcal{G}\} 
&\geq&
\log\left(P\Theta\right)-2\mathbb{E}\{\sqrt[3]{\log(\alpha_{\max}^{-1})}|\mathcal{G}\}
  \nonumber \\ & & \sqrt[3]{\log\log\left(P\Theta\right)} \left[1+
O\left(\frac{\log\log\log(\Theta)}{\log\log(\Theta)}\right)\right] -o(1)\notag\\
&\stackrel{a}{\geq}&
\log\left(P\Theta\right)-2\mathbb{E}\{\sqrt[3]{\log(\alpha_{\max}^{-1})}\}
  \nonumber \\ & & \sqrt[3]{\log\log\left(P\Theta\right)} \left[1+
O\left(\frac{\log\log\log(\Theta)}{\log\log(\Theta)}\right)\right] -o(1)\notag\\
&\stackrel{b}{\geq}&\log\left(P\Theta\right)-2\sqrt[3]{\mathbb{E}\{\log(\alpha_{\max}^{-1})\}}
 \nonumber \\ & & \sqrt[3]{\log\log\left(P\Theta\right)} \left[1+
O\left(\frac{\log\log\log(\Theta)}{\log\log(\Theta)}\right)\right]-o(1),
\end{eqnarray}
where ($a$) follows from the fact that $\mathbb{E}\{\sqrt[3]{\log(\alpha_{\max}^{-1})}|\mathcal{G}\} \leq \mathbb{E}\{\sqrt[3]{\log(\alpha_{\max}^{-1})}\} $ and ($b$) results from the convexity of cube root.
For large values of $K$,
$\mathbb{E}\{\log(\alpha_{\max}^{-1})\}$ can be approximated
 as follows (See Appendix B):
\begin{eqnarray}\label{kasy}
\mathbb{E}\{\log(\alpha_{\max}^{-1})\} &\backsimeq&
\frac{1}{Ke^{-\Theta}}
\left(1+O\left(\frac{1}{Ke^{-\Theta}}\right)\right)+
e^{-Ke^{-\Theta}} (\Theta - \log K).
\end{eqnarray}
Noting  (\ref{kasy}), for values of $\Theta$
satisfying (\ref{tetta}), we have
\begin{equation}\label{o1}
\mathbb{E}\{\log(\alpha_{\max}^{-1})\}\log\log\left(\Theta\right) \sim o(1).
\end{equation}
Using (\ref{Teq}) and (\ref{o1}), we can write
\begin{equation}\label{t3}
\mathbb{E}\{\mathcal{T}_3(\upsilon,\alpha_{\max})|\mathcal{G}\} \geq  \log\left(P\log K\right)- o(1).
\end{equation}
To compute $\mbox{Pr}\{\mathcal{G}\}$ defined in (\ref{gevent}), we
use Chebychev inequality.
\begin{eqnarray}\label{cheb}
\mbox{Pr}\left\{|\mathcal{Z}-\mathbb{E}\{\mathcal{Z}\}|
<\sqrt{\sqrt[3]{\log
K}\mbox{var}\{\mathcal{Z}\}}\right\}>1-\frac{1}{\sqrt[3]{\log K}},
\end{eqnarray}
where $\mathcal{Z}=N^{\rm{opt}}\log(\alpha_{\max}^{-1})$. Noting
(\ref{noptscal2}), (\ref{tetta}) and (\ref{kasy}), we have
\begin{eqnarray}\label{meanna}
\mathbb{E}\{N^{\rm{opt}}\log(\alpha_{\max}^{-1})\} &\stackrel{(\ref{noptscal2})}{\leq}&
\mathbb{E}\{\sqrt[3]{\log(\alpha_{\max}^{-1})\log\log\left(P\upsilon_{\max}\right)}\}\notag
\\ & = & \mathbb{E}\{\sqrt[3]{\log(\alpha_{\max}^{-1})}\}\mathbb{E}\{\sqrt[3]{\log\log\left(P\upsilon_{\max}\right)}\}\notag
\\ &\stackrel{a}{\lesssim} & \sqrt[3]{\mathbb{E}\{\log(\alpha_{\max}^{-1})\}}\sqrt[3]{\log\log\left(P\log K\right)} \notag \\
&\stackrel{(\ref{o1})}{=}&o(1)
\end{eqnarray}
where ($a$) follows from the fact the $\upsilon_{\max} \sim \log(K)$ with probability one.
Also, noting (\ref{noptscal2}), (\ref{tetta}) and (\ref{kasy}), we have
\begin{eqnarray}\label{varasy}
\mbox{var}\{N^{\rm{opt}}\log(\alpha_{\max}^{-1})\}&=&
\mbox{var}\{\sqrt[3]{\log(\alpha_{\max}^{-1})\log\log\left(P\upsilon\right)}\}\notag
\\&\leq&
\mathbb{E}\{(\log(\alpha_{\max}^{-1})\log\log\left(P\upsilon\right))^{\frac{2}{3}}\}\notag
\\ &\leq &
(\mathbb{E}\{{\log(\alpha_{\max}^{-1})}\})^{\frac{2}{3}} (\mathbb{E}\{\log\log\left(P\upsilon\right)\})^{\frac{2}{3}} \notag
\\ &\stackrel{(\ref{tetta}), a}{\lesssim} & O\left(\frac{1}{(\log K)^{\frac{2}{3}}}\right) O((\log \log \log K)^{\frac{2}{3}}) \notag
\\ & = & O\left(\left(\frac{\log \log \log K}{\log K}\right)^{\frac{2}{3}}\right)
\end{eqnarray}
where ($a$) follows from the fact that $\upsilon \sim \log K +O(\log
\log K)$ with probability one. Substituting (\ref{meanna}) and
(\ref{varasy}) in (\ref{cheb}), we have
\begin{eqnarray}\label{pr}
\mbox{Pr}\left\{|N^{\rm{opt}}\log(\alpha_{\max}^{-1})-o(1)| <
O\left(\left(\frac{\log \log\log K}{\sqrt{\log K}}\right)^{\frac{1}{3}}\right)\right\}>1- \frac{1}{\sqrt[3]{\log K}}
\end{eqnarray}
Noting (\ref{t31}), (\ref{t3}), and (\ref{pr}), the result of the theorem follows.
\rightline{$\blacksquare$}
\textit{Remark 1-} The uniform
distribution of the correlation coefficients is not a necessary
condition for Theorem \ref{s3}. In fact,  Theorem \ref{s3} is valid
if $\mbox{Pr}\{\mathcal{G}\}\sim 1- o\left(\frac{1}{\log \log K}\right)$. $\mbox{Pr}\{\mathcal{G}\}$ can be written as 
\begin{eqnarray}\label{rem}
\mbox{Pr}\{N^{\rm{opt}}\log(\alpha_{\max}^{-1})<
g(K)\}&=&\mbox{Pr}\{\sqrt[3]{\log(\alpha_{\max}^{-1})\log\log\left(P\log
K\right)}< g(K)\} \notag \\
&=&\mbox{Pr}\{\alpha_{\max}>e^{\frac{-{g(K)}^3}{\log\log\left(P\log
K\right)}}\} \notag \\
&=& 1- (F_{\alpha}(e^{\frac{-{g(K)}^3}{\log\log\left(P\log
K\right)}}))^K
\end{eqnarray}
where $g(K)$  satisfies $g(K)\sim o(1)$. Noting (\ref{rem}), there must exist a function $g(K)$ such that 
$F_{\alpha}(e^{\frac{-{g(K)}^3}{\log\log\left(P\log K\right)}})\sim
1-\omega\left(\frac{\log \log \log K}{K}\right)$ to satisfy $\mbox{Pr}\{\mathcal{G}\}\sim  1- o\left(\frac{1}{\log \log K}\right)$.
Hence, there exists a larger class of distributions that satisfy the
requirements for this theorem.

\section{Conclusion}\label{conclusion}
A multiuser downlink communication over a time-correlated fading
channel has been considered. We have proposed three scheduling
schemes in order to maximize the throughput of the system. Assuming
a large number of users in the system, we show that using SNR-based
scheduling, a gap of $\Omega(\sqrt{\log \log \log K})$ exists
between the achievable throughput and the maximum throughput of the
system. We propose a simple scheduling, considering both the SNR and
channel correlation of the users. We show that the gap between the throughput of the
proposed scheme and the maximum throughput of the system approaches zero as the
 number of users tends to infinity.

\section*{Appendix A}
For simplicity, we drop the user index. Noting (\ref{ekro}), we have
$E_0(\rho)= -\frac{1}{N}\log I_N$, where
\begin{equation}
I_N=\int_{u_N}...\int_{u_1} \prod_{i=1}^N
\left(\frac{1}{1+\frac{P}{1+\rho}u_i^2}\right)^{\rho} p(\textbf{u}|u_0) du_i.
\end{equation}
Using (\ref{corrfading}), we have
\begin{eqnarray}
I_N=\int_{u_N}...\int_{u_1} \prod_{i=1}^N \frac{2u_i}{1-\alpha^2}\exp\left\{-\frac{u_i^2+\alpha^2u_{i-1}^2}{1-\alpha^2}\right\}
 \mathcal{I}_0\left(\frac{2\alpha u_i u_{i-1}}{1-\alpha^2}\right)\left(\frac{1}{1+\frac{P}{1+\rho}u_i^2}\right)^{\rho }  du_i.
\end{eqnarray}
Substituting $v_i=\frac{u_i}{u_0\sqrt{(1-\alpha^2)/2}}$,  $0\leq i \leq N$,  we have
\begin{eqnarray}\label{IN}
I_N=\int_{v_N}...\int_{v_1} \prod_{i=1}^N u_0^2v_i e^{-\frac{v_i^2+\alpha^2v_{i-1}^2}{2/u_0^2}} \mathcal{I}_0(\alpha u_0^2 v_i v_{i-1})  f(v_i)  dv_i \nonumber \\
=\int_{v_N}...\int_{v_1} \prod_{i=1}^N u_0^2v_i e^{-\frac{(v_i -\alpha v_{i-1})^2}{2/u_0^2}}e^{-\alpha u_0^2 v_i v_{i-1}} \mathcal{I}_0(\alpha u_0^2 v_i v_{i-1})f(v_i)  dv_i,
\end{eqnarray}
where,
\begin{equation}\label{fdef}
f(v_i)=\left(\frac{1}{1+\frac{Pu_0^2(1-\alpha^2)}{2(1+\rho)}v_i^2}\right)^{\rho}.
\end{equation}
For large values of $u_0$, we evaluate the following integral.
\begin{eqnarray}
I&=& u_0^2 v e^{-\frac{(v -\mu)^2}{2/u_0^2}}e^{-u_0^2 v \mu} \mathcal{I}_0( u_0^2 v \mu)\varphi(v) \notag \\
&=&\int_{0}^{\infty} g(v) \frac{1}{\sqrt{2\pi/u_0^2}}e^{-\frac{(v-\mu)^2}{2/u_0^2}} dv,
\end{eqnarray}
where $g(v) \triangleq \sqrt{2\pi} v u_0 \mathcal{I}_0( u_0^2 v \mu)e^{- u_0^2 v \mu }\varphi(v)$ and $\varphi(v)$ is differentiable and satisfies $0 \leq \varphi(v) \leq 1$ and $\varphi(v)\sim O(\frac{1}{u_0^\rho})$. Noting that ~\cite{approx}
\begin{eqnarray}\label{beselaprox}
\mathcal{I}_0(z)e^{-z}\sqrt{2\pi z}=1+O\left(\frac{1}{z}\right),  \qquad z\gg 1 
\end{eqnarray}
it is easy to show that $g^{(n)}(\mu)$ is bounded for $\mu \geq 0$ and $n\geq 1$. Using Taylor series of $g(\upsilon)$ about $\mu$, we have
\begin{eqnarray}\label{Ig}
I&=&\int_{0}^{\infty} \left(g(\mu)+ \sum_{n=1}^{\infty} \frac{g^{(n)}(\mu)}{n!}(v-\mu)^n \right) \frac{1}{\sqrt{2\pi/u_0^2}}e^{-\frac{(v-\mu)^2}{2/u_0^2}} dv \notag \\
&=& g(\mu)(1-Q(\mu u_0))+ \int_{0}^{\infty} \sum_{n=1}^{\infty} \frac{g^{(n)}(\mu)}{n!}(v-\mu)^n \frac{1}{\sqrt{2\pi/u_0^2}}e^{-\frac{(v-\mu)^2}{2/u_0^2}} dv  \notag \\
&=& g(\mu)(1-Q(\mu u_0))+ \int_{[\mu-\frac{1}{\sqrt{u_0}}]^+}^{\mu+\frac{1}{\sqrt{u_0}}} \sum_{n=1}^{\infty}\frac{g^{(n)}(\mu)}{n!}(v-\mu)^n \frac{1}{\sqrt{2\pi/u_0^2}}e^{-\frac{(v-\mu)^2}{2/u_0^2}} dv +\varepsilon \notag \\
&=& g(\mu)(1-Q(\mu u_0)) + O\left(\frac{g'(\mu)}{\sqrt{u_0}}\right)+\varepsilon.
\end{eqnarray}
where  $\varepsilon$ can be bounded as follows:
\begin{eqnarray}\label{epsil}
\varepsilon &\stackrel{a}{\leq}& \int_{0}^{[\mu-\frac{1}{\sqrt{u_0}}]^+} \frac{g(v)}{\sqrt{2\pi/u_0^2}}e^{-\frac{(v-\mu)^2}{2/u_0^2}} dv +\int_{\mu+\frac{1}{\sqrt{u_0}}}^{\infty} \frac{g(v)}{\sqrt{2\pi/u_0^2}}e^{-\frac{(v-\mu)^2}{2/u_0^2}} dv \nonumber \\
 & \stackrel{b}{\leq} & \sqrt{2\pi}u_0 \int_{0}^{[\mu-\frac{1}{\sqrt{u_0}}]^+}   \frac{v}{\sqrt{2\pi/u_0^2}} e^{-\frac{(v-\mu)^2}{2/u_0^2}} dv +  \sqrt{2\pi}u_0 \int_{\mu+\frac{1}{\sqrt{u_0}}}^{\infty} \frac{v}{\sqrt{2\pi/u_0^2}} e^{-\frac{(v-\mu)^2}{2/u_0^2}} dv \nonumber \\
&\leq& 2\sqrt{2\pi}u_0 \int_{\mu+\frac{1}{\sqrt{u_0}}}^{\infty} \frac{v}{\sqrt{2\pi/u_0^2}} e^{-\frac{(v-\mu)^2}{2/u_0^2}} dv \nonumber \\
 &=&  2\sqrt{2\pi}u_0 \left(Q(\sqrt{u_0})\left(\mu+\frac{1}{\sqrt{u_0}}\right)+ \int_{\sqrt{u_0}}^{\infty} Q(z) dz \right)\notag \\
&\stackrel{c}{\leq}&  2\sqrt{2\pi}u_0 \left(\left(\mu+\frac{1}{\sqrt{u_0}}\right)e^{-\frac{u_0}{2}} + \int_{\sqrt{u_0}}^{\infty} e^{-\frac{z^2}{2}} dz \right)\notag \\
&\stackrel{d}{\leq}& 2\sqrt{2\pi} e^{-\frac{u_0}{2}} u_0 \left(\left(\mu+\frac{1}{\sqrt{u_0}}\right) + \sqrt{2\pi} \right) \notag \\
&\leq& O\left(u_0e^{-\frac{u_0}{2}}\right)
\end{eqnarray}
where ($a$) results from the fact that $g(\mu)\geq 0$, ($b$) is valid because $\mathcal{I}_0(\mu z)e^{-\mu z}\leq  1$ for $ \mu\geq 0$ and $z\geq 0$,  and ($c$) and ($d$) follow from the fact that $Q(z)\triangleq \frac{1}{\sqrt{2\pi}}\int_z^{\infty} e^{-t^2/2} dt \leq e^{-z^2/2}$.
Moreover, using (\ref{beselaprox}), we can write
\begin{eqnarray}\label{fg}
g(\mu)&=& \varphi(\mu) \sqrt{2\pi} \mu u_0 \mathcal{I}_0( u_0^2 \mu^2)e^{- u_0^2 \mu^2 } \notag \\
& =&\varphi(\mu)\left(1+O\left(\frac{1}{u_0^2}\right)\right).
\end{eqnarray}
Also, using (\ref{beselaprox}) and noting $\varphi(v)\sim O(\frac{1}{u_0^\rho})$, we have
\begin{eqnarray}\label{gg}
g(v)= \varphi(v)\sqrt{\frac{v}{\mu}}\left(1+O\left(\frac{1}{u_0^2}\right)\right)\notag \\
\Rightarrow O(g'(v))=O\left(\frac{\varphi(v)}{2\sqrt{v\mu}}+\varphi'(v)\sqrt{\frac{v}{\mu}}\right) \notag \\
\Rightarrow O(g'(\mu))=O(\varphi(\mu)).
\end{eqnarray}
Using (\ref{Ig}), (\ref{epsil}), (\ref{fg}) and (\ref{gg}), we have
\begin{eqnarray}\label{If}
I&=&\varphi(\mu)\left(1+O\left(\frac{1}{\sqrt{u_0}}\right)\right)+O\left(u_0e^{-\frac{u_0}{2}}\right)\notag \\
&\stackrel{a}{=}&\varphi(\mu)\left(1+O\left(\frac{1}{\sqrt{u_0}}\right)\right),
\end{eqnarray}
where ($a$) follows from the fact that $\varphi(\mu)=O\left(\frac{1}{u_0^{\rho}}\right)$.
Applying (\ref{If}) in (\ref{IN}), we have
\begin{eqnarray}\label{In}
I_N &=&\int_{v_{N-1}}...\int_{v_1} f(\alpha v_{N-1})\left(1+O\left(\frac{1}{\sqrt{u_0}}\right)\right)\left( 1-Q(\alpha v_{N-1} u_0) \right)\times \nonumber \\ & & \prod_{i=1}^{N-1} u_0^2v_i e^{-\frac{(v_i -\alpha v_{i-1})^2}{2/u_0^2}-\alpha u_0^2 v_i v_{i-1}}\mathcal{I}_0(\alpha u_0^2 v_i v_{i-1})f(v_i)  dv_i \nonumber \\
&=&\int_{v_{N-2}}...\int_{v_1} f(\alpha^2 v_{N-2})f(\alpha v_{N-2})\left(1+O\left(\frac{1}{\sqrt{u_0}}\right)\right)^2\left( 1-Q(\alpha^2 v_{N-2} u_0) \right)\times \nonumber \\ & &\left( 1-Q(\alpha v_{N-2} u_0) \right) \prod_{i=1}^{N-2} u_0^2v_i  e^{-\frac{(v_i -\alpha v_{i-1})^2}{2/u_0^2}-\alpha u_0^2 v_i v_{i-1}}\mathcal{I}_0(\alpha u_0^2 v_i v_{i-1})f(v_i)  dv_i \nonumber \\& =& \cdots =\prod_{i=1}^N f(\alpha^i v_0)\left(1+O\left(\frac{1}{\sqrt{u_0}}\right)\right)\left( 1-Q(\alpha^i v_0 u_0) \right)
\end{eqnarray}
Substituting  $v_0=\frac{1}{\sqrt{(1-\alpha^2)/2}}$, we have
\begin{eqnarray}\label{inapp}
I_N &=& \prod_{i=1}^N f\left(\frac{\sqrt{2}\alpha^i}{\sqrt{(1-\alpha^2)}}\right)\left(1+O\left(\frac{1}{\sqrt{u_0}}\right)\right)\left( 1-Q\left(\frac{\sqrt{2}\alpha^i u_0}{\sqrt{(1-\alpha^2)}}\right) \right) \notag \\
&=&\prod_{i=1}^N f\left(\frac{\sqrt{2}\alpha^i}{\sqrt{(1-\alpha^2)}}\right)\left(1+O\left(\frac{1}{\sqrt{u_0}}\right)\right)\left( 1-O\left(e^{-u^2_0}\right) \right) 
\end{eqnarray}
Using (\ref{fdef}) and (\ref{inapp}) and noting  $E_0(\rho)=
-\frac{1}{N}\log I_N$, we conclude Theorem \ref{throu}.

\section{Appendix B}
$\mathbb{E}\{\log(\alpha_{max}^{-1})\}$ can be derived as follows
\begin{eqnarray} \label{pa}
\mathbb{E}\{\log(\alpha_{\max}^{-1})\} = \sum_{n=1}^{K} \left.
\mathbb{E}\{\log(\alpha_{\max}^{-1})\right| |{\cal S}|=n\} \mbox{Pr}
\{ |{\cal S}|=n\}.
\end{eqnarray}
Since $\alpha_k$, $k=1, \cdots, K$, are i.i.d. random variables with
uniform distribution, we can write
\begin{eqnarray} \label{pa0}
F_{\alpha_{\max}} (\alpha \left| |{\cal S}|=n  )\right. &=& \alpha^n  \notag\\
\Rightarrow \left. \mathbb{E} \{ \log(\alpha_{\max}^{-1})\right|
|{\cal S}|=n\} &=& \int_{0}^{1} \log (\alpha^{-1}) n \alpha^{n-1}
d \alpha \notag\\
&=& \frac{1}{n},
\end{eqnarray}
where $F_{X} (.)$ denotes the cumulative density function  of the
random variable $X$. Indeed, $|{\cal S}|$ is a binomial random
variable with parameters $K$ and $e^{-\Theta}$. (Since $\upsilon_k=u_{0,k}^2$ and $u_{0,k}$  has a Rayleigh distribution, we have
$\mbox{Pr}(\upsilon_k \geq \Theta)=e^{-\Theta}$). Hence,
 \begin{eqnarray}\label{pa1}
\mbox{Pr}\{|{\cal S}|=n\}
=\binom{K}{n}e^{-n\Theta}(1-e^{-\Theta})^{K-n}.
\end{eqnarray}
Substituting (\ref{pa0}) and (\ref{pa1}) in (\ref{pa}), we have
\begin{eqnarray}\label{ealpha0}
\mathbb{E}\{\log(\alpha_{\max}^{-1})\}  =  \sum_{n=1}^{K}
\binom{K}{n} \frac{1}{n} e^{-n\Theta} (1-e^{-\Theta})^{K-n}.
\end{eqnarray}
Let us define $\lambda(K)\triangleq \sum_{n=1}^{K} \binom{K}{n}\frac{1}{n}
e^{-n\Theta}(1-e^{-\Theta})^{K-n}$.
\begin{eqnarray}
\lambda(K)=\sum_{n=0}^{K-1} \binom{K}{n+1}\frac{e^{-(n+1)\Theta}}{n+1} (1-e^{-\Theta})^{K-n-1} \nonumber \\
=(1-e^{-\Theta})\sum_{n=0}^{K-2} \binom{K-1}{n+1}\frac{e^{-(n+1)\Theta}}{n+1} (1-e^{-\Theta})^{K-n-2} +\nonumber \\
e^{-\Theta}\sum_{n=0}^{K-1} \binom{K-1}{n}\frac{1}{n+1} e^{-n\Theta}(1-e^{-\Theta})^{K-n-1} \nonumber \\
=(1-e^{-\Theta})\lambda(K-1)+\frac{1}{K}-\frac{(1-e^{-\Theta})^K}{K}
\end{eqnarray}
Solving the iteration considering $\lambda(1)=e^{-\Theta}$, we derive
$\lambda(K)$ which is equal to $\mathbb{E}\{\log(\alpha_{\max}^{-1})\}$.
\begin{equation} \label{ealpha}
\mathbb{E}\{\log(\alpha_{\max}^{-1})\}=\sum_{n=1}^{K}
\frac{1}{n}(1-e^{-\Theta})^{K-n}-(1-e^{-\Theta})^{K}\sum_{n=1}^{K}\frac{1}{n}.
\end{equation}
For large values of $K$, we can approximate (\ref{ealpha}) as
\begin{eqnarray}\label{kasyB}
\mathbb{E}\{\log(\alpha_{\max}^{-1})\} &\backsimeq&
(1-e^{-\Theta})^K \left(\int_{1}^{K} \frac{(1-e^{-\Theta})^{-x}dx}{x} -\int_{1}^{K}\frac{dx}{x} \right){} \nonumber \\
&\stackrel{a}{\backsimeq}&\frac{1}{-K\log(1-e^{-\Theta})}\left(1+O\left(\frac{1}{-K\log(1-e^{-\Theta})}\right)\right)
\nonumber \\ & & +(1-e^{-\Theta})^K (\Theta-\log K) \notag\\
&\backsimeq& \frac{1}{Ke^{-\Theta}}
\left(1+O\left(\frac{1}{Ke^{-\Theta}}\right)\right)+
e^{-Ke^{-\Theta}} (\Theta - \log K).
\end{eqnarray}
where ($a$) results from the following approximations~\cite{approx}:
\begin{eqnarray}
\int_{-\tau}^{\infty} \frac{e^{-t}}{t} dt \simeq \frac{e^{\tau}}{\tau}\left(1+O\left(\frac{1}{\tau}\right)\right)-i\pi   \qquad \tau \gg 1 \\
\int_{-\tau}^{\infty} \frac{e^{-t}}{t} dt \simeq \log(\tau) -i\pi   \qquad  0 <\tau \ll 1
\end{eqnarray}

\end{document}